\documentclass[12pt,preprint]{aastex}
\begin{document}

\title{MMT Survey for Intervening \ion{Mg}{2}
Absorption\altaffilmark{1}}

\author{Daniel B. Nestor\altaffilmark{2,3}, David
 A. Turnshek\altaffilmark{2}, and Sandhya M. Rao\altaffilmark{2}}

\altaffiltext{1}{Observations reported here were obtained at the MMT
Observatory, a joint facility of the Smithsonian Institution and the
University of Arizona.}

\altaffiltext{2}{Department of Physics \& Astronomy, University of
Pittsburgh, Pittsburgh, PA 15260; turnshek@quasar.phyast.pitt.edu;
rao@everest.phyast.pitt.edu}

\altaffiltext{3}{also Department of Astronomy, University of Florida,
Gainesville, FL 32611; dbn@astro.ufl.edu}

\begin{abstract}
We present the results from a spectroscopic survey for intervening
\ion{Mg}{2} absorption in the spectra of 381 background QSOs conducted
at the Multiple Mirror Telescope.  This survey complements our earlier
SDSS EDR \ion{Mg}{2} survey, extending our results to lower redshift
($z \simeq 0.15$) and weaker \ion{Mg}{2} $\lambda2796$ rest equivalent
width ($W_0^{\lambda2796} \simeq 0.1$\AA).  We confirm two major
results from that survey: the transition in the $W_0^{\lambda2796}$
distribution at $W_0^{\lambda2796} \approx 0.3$\AA, and the
$W_0^{\lambda2796}$-dependent evolution of the incidence of systems.
The nature of $\partial^2N/\partial z \partial W_0^{\lambda2796}$ is
consistent with the idea that multiple physically-distinct
components/processes contribute to the incidence of \ion{Mg}{2}
absorption systems in a $W_0$-dependent manner and evolve at different
rates.  A significant decrease in the total proper absorption cross
section is detected in our MMT data for systems as weak as 1.0 \AA\
$\le W_0^{\lambda2796} < 1.5$\AA\ at $z\lesssim 0.4$.  We discuss this
$W_0$-dependent evolution in the context of the evolution of galaxy
structures, processes including superwinds and interactions,  and
damped-Ly$\alpha$ absorbers.  We also consider the possibility that
the observed redshift and $W_0^{\lambda2796}$ dependence of the
incidence of absorption in spectroscopic surveys for low-ion/neutral
gas results from the effects of dust-induced extinction.

\end{abstract}


\section{Introduction}
Over the past 25 years, spectroscopic QSO surveys have established the
statistics of intervening metal-line absorption systems, most notably
those selected via \ion{C}{4} and \ion{Mg}{2} resonance transitions
(e.g., Weyman et al. 1979; Lanzetta, Turnshek, \& Wolfe 1987; Tytler
et al. 1987; Sargent, Boksenberg \& Steidel 1988; Steidel \& Sargent
1992, hereafter SS92; Churchill, Rigby, Charlton, \& Vogt 1999,
hereafter CRCV99; Nestor, Turnshek, \& Rao 2005a, hereafter NTR05).
These systems identify structures based on their gas absorption cross
section, and  are widely believed to select galactic structures and
extended gaseous galactic halos (e.g., Bergeron \& Boiss\'{e}, 1991;
Steidel et. al, 2002; Churchill, Kacprzak \& Steidel 2005).  The
results of the early \ion{Mg}{2} surveys, which we summarize in NTR05,
indicated that the distribution of rest equivalent widths of the
$\lambda2796$ line ($W_0^{\lambda2796}$) can be described by a power
law or an exponential, and that the incidence of the strongest systems
seems to be decreasing with decreasing redshift.  In NTR05 we
presented a large ($\approx 1,300$ systems) \ion{Mg}{2} survey using
the Sloan Digital Sky Survey (SDSS) Early Data Release (EDR) QSO
spectra.  We demonstrated that the $W_0^{\lambda2796}$ distribution is
fit very well by an exponential over the range 0.3\AA\ $\le
W_0^{\lambda2796} \lesssim 5$\AA\ and that previous power law fits
over predict the incidence of strong lines.  However, extrapolating
the NTR05 exponential to weak (0.02\AA\ $\le
W_0^{\lambda2796}<0.3$\AA) systems under predicts their incidence when
compared to the results of CRCV99.  This apparent transition is
consistent with the idea that weak \ion{Mg}{2} absorbers are in part
comprised of a population that  is physically distinct from
intermediate/strong \ion{Mg}{2} systems (e.g., Rigby, Charlton \&
Churchill 2002).  Since the transition occurs at the boundary of two
different surveys, however, it is important to investigate the
possibility that this is simply an artifact of some unknown systematic
difference between the surveys.

NTR05 also investigated the incidence of \ion{Mg}{2} systems as a
function of  redshift for different ranges of $W_0^{\lambda2796}$.
The results were consistent with the total proper cross section for
absorption decreasing with decreasing redshift, especially at
$z\lesssim1$, such that the stronger lines evolve faster.  The
apparent evolution can be described by a steepening of the
$W_0^{\lambda2796}$ distribution with cosmic time.  However, the data
only reach $z\ge0.37$ and the evolutionary signal has high
significance only for very strong ($W_0^{\lambda2796} \gtrsim 2.0$\AA)
systems.  Thus, extending the measurement of $\partial N/\partial z$
to lower redshift is a major goal of this work.  Furthermore, the
evolution of \ion{Mg}{2} systems over cosmic time has a more general
significance, since \ion{Mg}{2} absorbers trace neutral hydrogen.
Extending the statistics of \ion{Mg}{2} absorption to lower redshift
is therefore important for the study of the evolution of the
\ion{H}{1} content of the Universe as well.  In fact, results from
this survey have already been used to study the incidence of damped
Ly$\alpha$ systems at low redshift (Rao, Turnshek \& Nestor, 2005).

Accurate measurements of the incidence of \ion{Mg}{2} absorption
systems as a function of $W_0^{\lambda2796}$ and redshift for weak to
``ultra-strong''  (4\AA\ $\lesssim W_0^{\lambda2796}\lesssim6$\AA)
systems and from high to low redshift, together with kinematic and
relative abundance information from high-resolution spectroscopy (see,
e.g., Ding, Charlton, \& Churchill, 2005, and references therein) and
composite-spectrum analysis (e.g., Nestor, Rao, Turnshek \& Vanden
Berk, 2003), are allowing the ability to understand the physical
nature and evolution of the structures selected by such systems.  Here
we present the results of a spectroscopic QSO survey conducted at the
6.5m Multiple Mirror Telescope on Mount Hopkins, AZ.  The goals of the
survey are: 1) to provide a survey large enough to detect a
significant number of intermediate/strong systems (0.3\AA $\le
W_0^{\lambda2796} \lesssim 2$\AA) and sensitive enough to measure weak
(0.1\AA\ $\le W_0^{\lambda2796} < 0.3$\AA) systems in a single survey;
and 2) extend the measurement of their incidence to lower redshift
($z\approx 0.15$).  In \S2 we describe the observations and our data
reduction procedure.  We present the results in \S3.  In \S4, we
discuss the relevance of our results to the problem of understanding
the nature of low-ion metal-line absorbers, and present our
conclusions in \S5.

\section{Observations}
Observations took place over a 42 month period from January of 2001 to
May of 2004 at the 6.5m MMT on the summit of Mount Hopkins, AZ.
Useful data were collected on 17 nights spread over seven observing
runs.  The MMT spectrograph was operated using the blue-channel
optical layout with the 800 grooves/mm grating, which corresponds to
0.75\AA/pixel and a resolution of approximately 2.2\AA.  Exposure
times varied with the QSO magnitude and varying observing conditions,
but were typically ten to fifteen minutes per object.  Approximately
900 spectra of nearly 400 QSOs were collected.  Quartz lamp exposures
were used  to correct for pixel to pixel sensitivity variances, and
comparison lamp exposures (usually He-Ne-Ar) were used to wavelength
calibrate the spectra.  The wavelength coverage varied slightly for
each observing run, but typically spanned $\approx 3190$\AA\ to either
$\approx 5170$\AA\ or $\approx 5465$\AA.

The resulting data set comprised a total of 381 useful QSO sightlines.
Continua were fitted to the reduced QSO spectra and \ion{Mg}{2}
doublet candidates found, inspected, and measured in a manner similar
to that described in NTR05.  The same software were used, though with
slight modifications to account for the differences in resolution and
wavelength coverage between the SDSS and MMT data.

\section{Results}
\label{sec_results}
The final MMT-survey \ion{Mg}{2} absorber sample consists of a total
of 140 doublets.  They cover the redshift range of $0.1973 \le z \le
0.9265$ and have $W_0^{\lambda2796}$ values ranging from 0.128\AA\ to
3.165\AA.

\begin{figure}
\plottwo{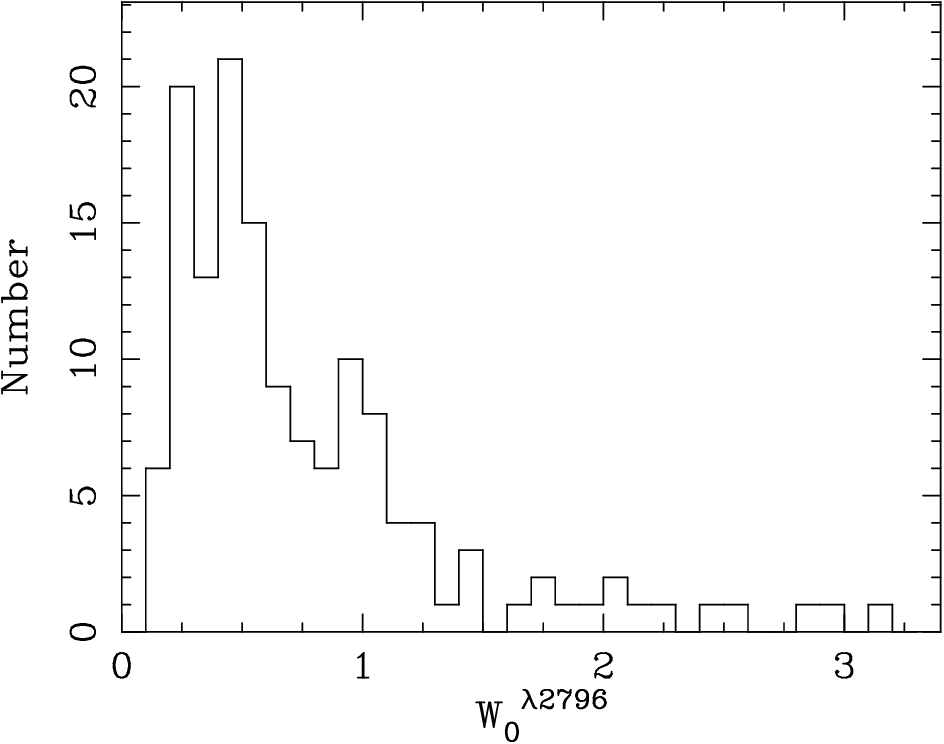}{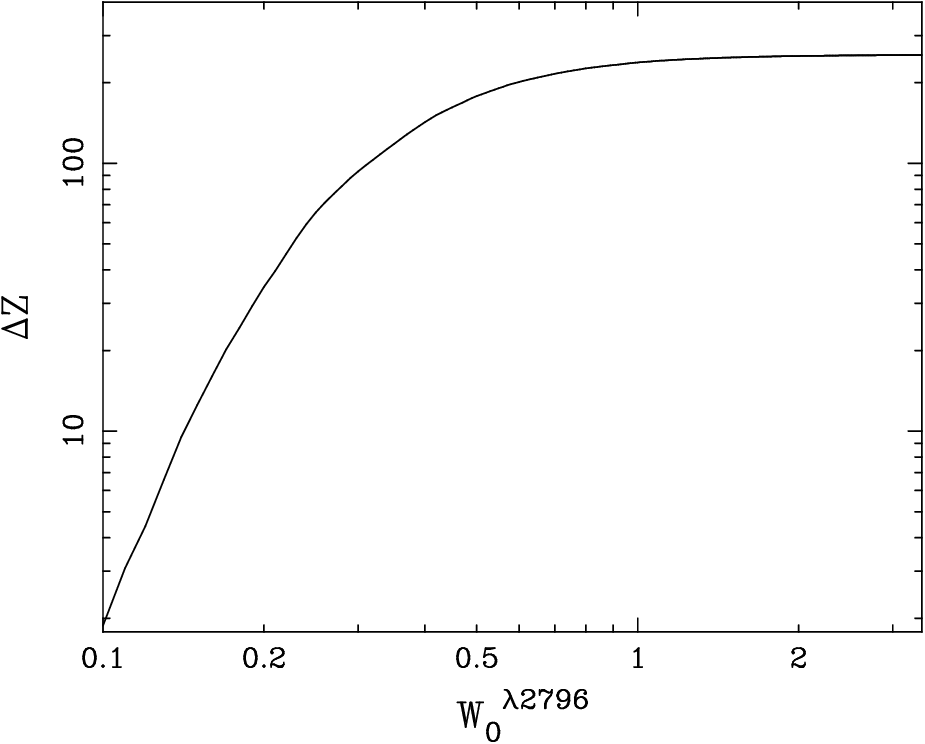}
\caption{The $W_0^{\lambda2796}$ distribution.  Left: The distribution
of rest equivalent widths, $W_0^{\lambda2796}$,  for \ion{Mg}{2}
systems found in the survey.  Right:  The redshift-path covered by the
survey, $\Delta Z\,(W_0^{\lambda2796}) =
\int_{z_{min}}^{z_{max}}\,\sum_i^{N_{spec}}\,g_i(W_0^{\lambda2796},z)\,dz$,
as a function of $W_0^{\lambda2796}$.}
\label{fig_mmtW}
\end{figure}

\begin{figure}
\plotone{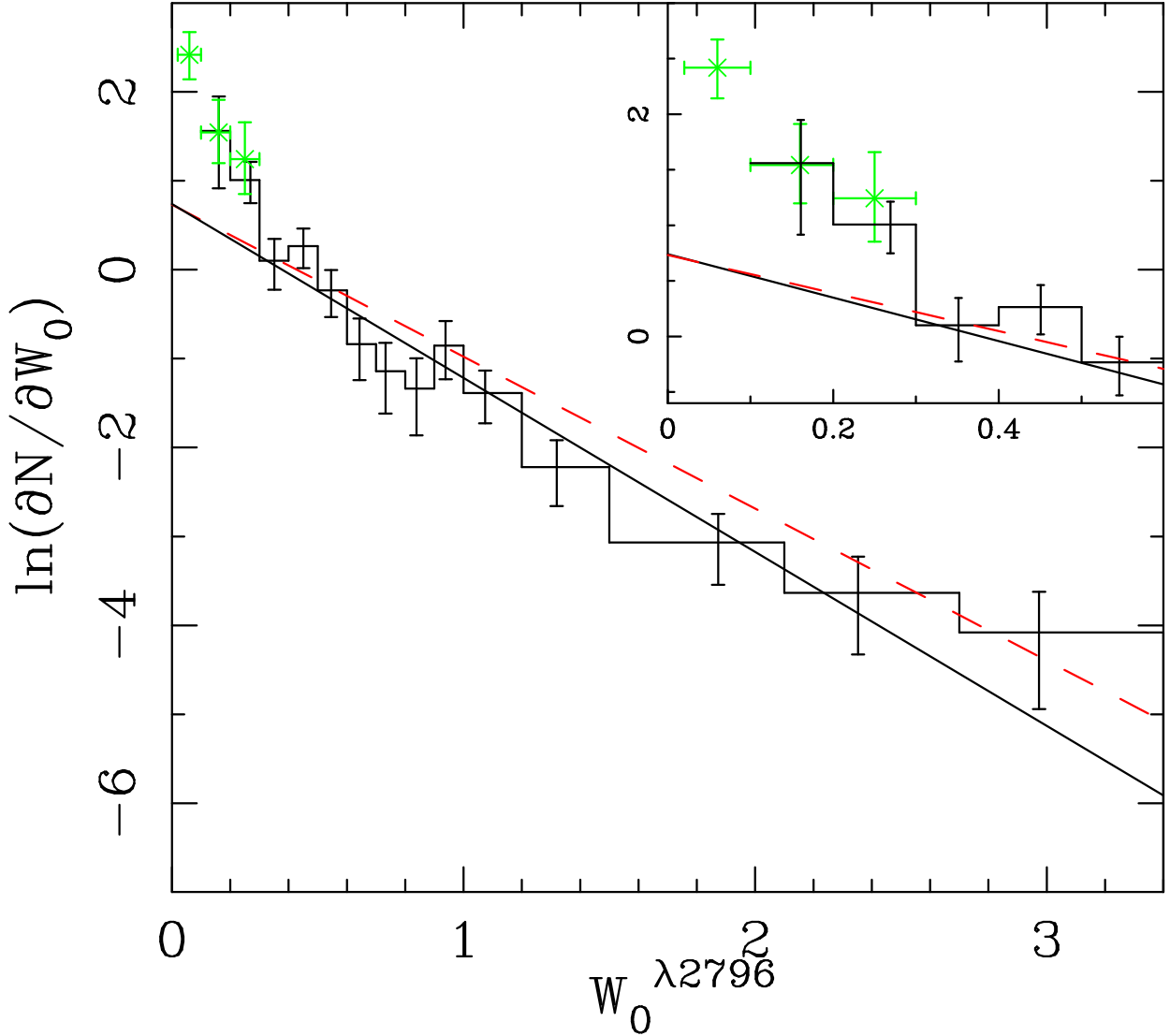}
\caption{The redshift-path corrected $W_0^{\lambda2796}$ distribution
from the MMT data.  The mean absorber redshift is
$\left<z\right>=0.589$.  The solid line is a maximum likelihood fit of
the form $\partial N/\partial W_0^{\lambda2796} = \frac{N^*}{W^*}
e^{-\frac{W_0}{W^*}}$ to data having $W_0^{\lambda2796}>0.3$\AA, with
$W^*=0.511 \pm 0.047$ and $N^*=1.071 \pm 0.119$.  The red-dashed line
is the low-$z$ result from the SDSS EDR survey with
$\left<z\right>=0.655$.  The points represent data from CRCV99 with
$\left<z\right>=0.9$.  The MMT result confirms the
upturn in $\partial N/\partial W_0^{\lambda2796}$ below 0.3\AA\ that
was first identified in NTR05.}
\label{fig_mmtrewd}
\end{figure}

\begin{figure}
\plotone{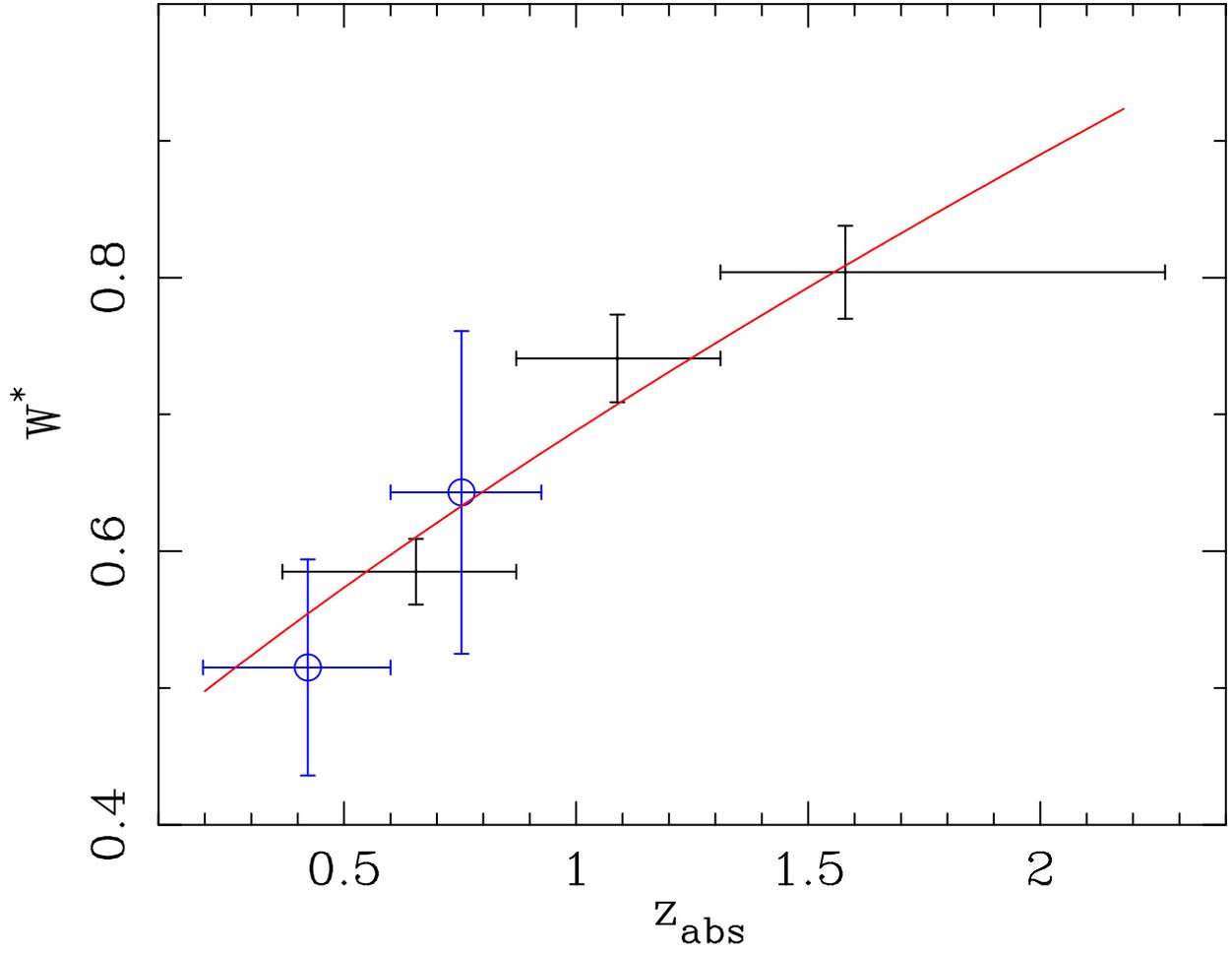}
\caption{Evolution of $W^*$.  The points without symbols are from the
EDR survey and the circles are the MMT result for
$W_0^{\lambda2796}>0.5$\AA.  The curve is the power law fit described
in NTR05.}
\label{fig_mmtdwdz}
\end{figure}

Figure \ref{fig_mmtW} shows the distribution of $W_0^{\lambda2796}$.
The redshift-path coverage as a function of $W_0^{\lambda2796}$ is
also shown.  The resulting $\partial N/\partial W_0^{\lambda2796}$ is
shown as the histogram in Figure \ref{fig_mmtrewd}.  The solid line is
a maximum likelihood fit to the data having
$W_0^{\lambda2796}\ge0.3$\AA\ of the form $\partial N/\partial
W_0^{\lambda2796} = \frac{N^*}{W^*} e^{-\frac{W_0}{W^*}}$ (see NTR05)
with $W^*=0.509 \pm 0.047$ and $N^*=1.089 \pm 0.121$.  The red-dashed
line is the fit to the low-$z$ data from the SDSS EDR survey, with a
mean redshift $\left<z_{abs}\right>=0.655$.  The systems found in the
MMT survey have $\left<z_{abs}\right>=0.589$.  The points represent
data from CRCV99, with $\left<z_{abs}\right>=0.9$.  The distribution
determined from the MMT data is consistent with both the SDSS EDR and
the CRCV99 results.  Although there are only 26 systems in the two
lowest $W_0^{\lambda2796}$ bins of the MMT survey, $\partial
N/\partial W_0^{\lambda2796}$ for $W_0^{\lambda2796}<0.3$\AA\ is in
very good agreement with the CRCV99 results and significantly
($1.4\sigma$ and $2.4\sigma$) above the extrapolation of the
single-exponent fit to $\partial N/\partial W_0^{\lambda2796}$ for
$W_0^{\lambda2796} \ge 0.3$\AA.  Though the CRCV99 data has a higher
mean redshift, weak \ion{Mg}{2} systems exhibit the least evolution.
Correcting the $W_0^{\lambda2796} < 0.3$\AA\ points for evolution has
only a minimal effect on Figure \ref{fig_mmtrewd} and makes the
consistency more robust.  Thus, the results from the MMT survey
confirm the upturn in $\partial N/\partial W_0^{\lambda2796}$ below
0.3\AA\ that was first identified in NTR05.  Figure \ref{fig_mmtdwdz}
shows $W^*_{\mathrm{MMT}}$ determined for systems with
$W_0^{\lambda2796} \ge 0.5$\AA, along with the values from the SDSS
EDR survey.  The $W_0^{\lambda2796}$ distribution determined from the
MMT survey is consistent within the errors with the SDSS EDR survey
results, considering the observed redshift evolution.

\begin{figure}
\plottwo{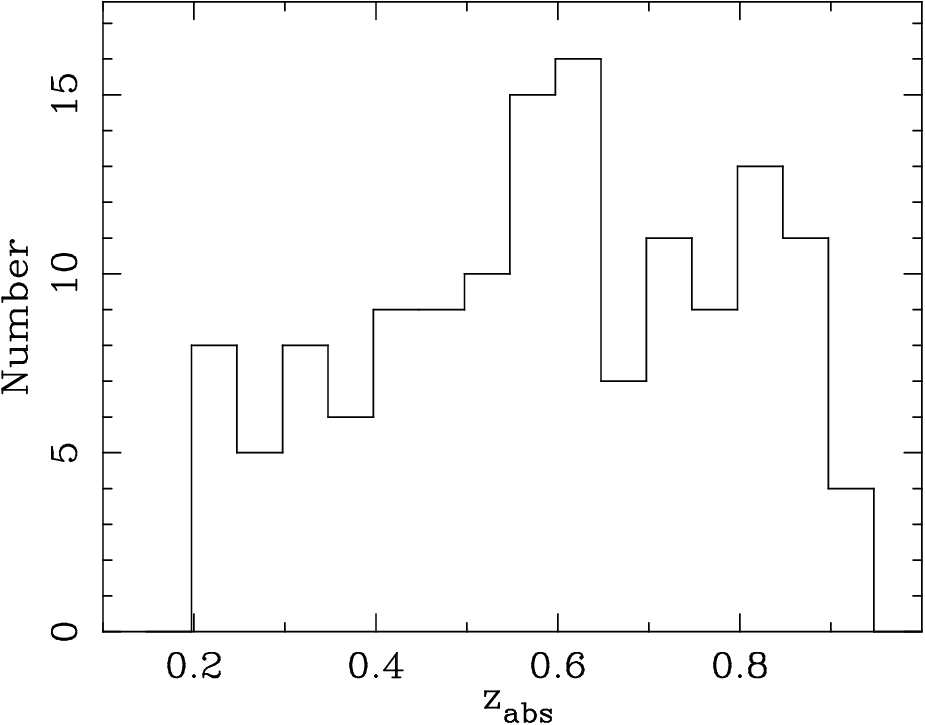}{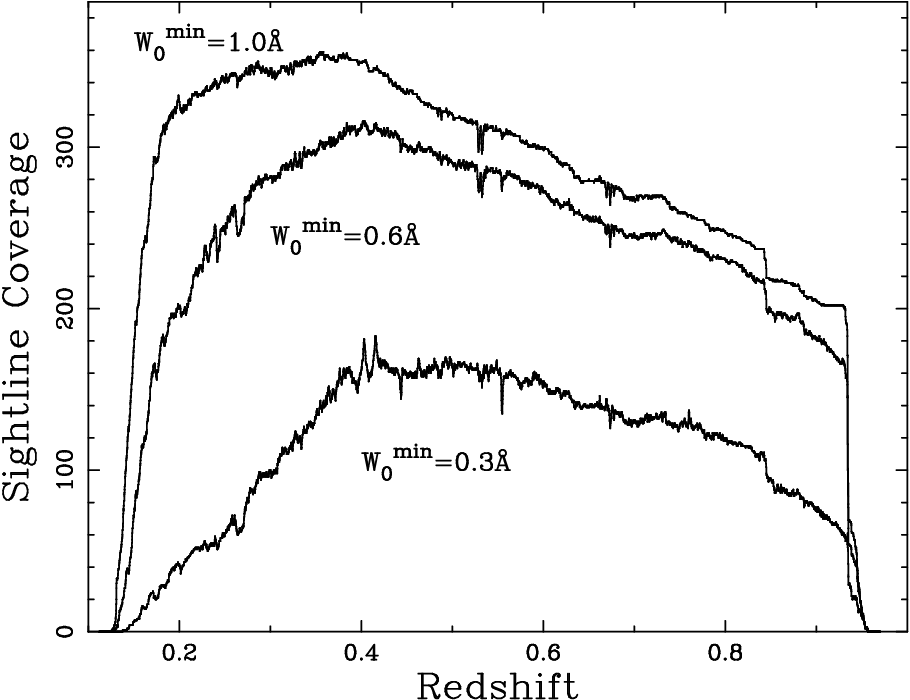}
\caption{The absorption redshift distribution.  Left: The distribution
of redshifts for \ion{Mg}{2} absorption systems found in the MMT
survey.  Right: The total number of  sightlines with sufficient signal
to noise ratio to detect lines with  $W_0^{\lambda2796} \ge W_0^{min}$
as a function of redshift, for $W_0^{min}$ = 1.0\AA, 0.6\AA, and
0.3\AA.}
\label{fig_mmtz}
\end{figure}

\begin{figure}
\plotone{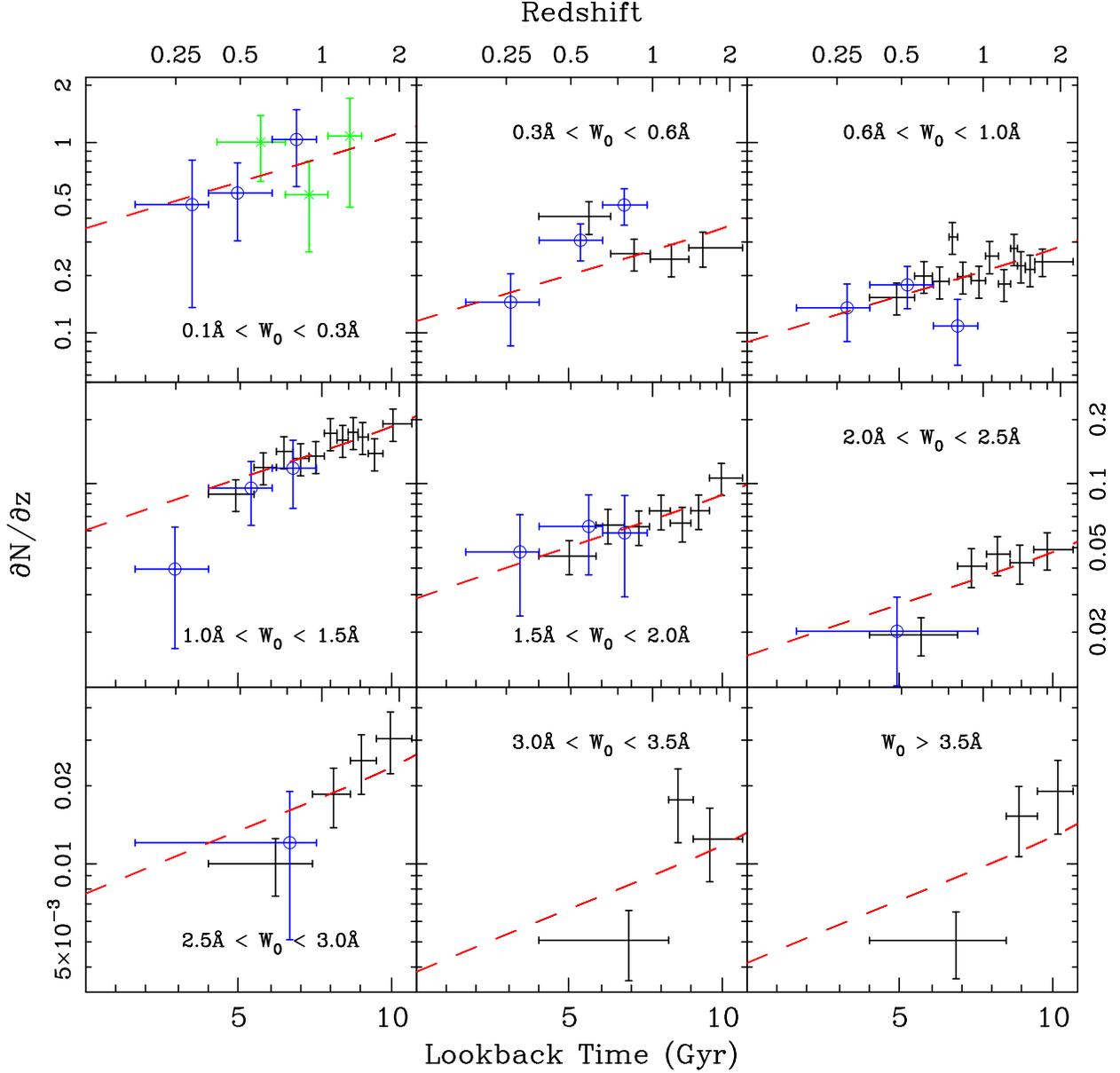}
\caption{$\partial N/\partial z$ as a function of look-back time, for
a $(\Omega_\lambda,\Omega_M,h)=(0.7,0.3,0.7)$ cosmology.  The
horizontal bars represent bin size and the vertical bars the 1$\sigma$
uncertainties.  Points without symbols are the EDR results, while the
circles represent the MMT and $\times$ symbols the CRCV99 results.
The dashed lines are the no-evolution curves normalized to the 
binned EDR/CRCV99 data, but excluding the MMT data.}
\label{fig_mmtdndz}
\end{figure}

The distribution of redshifts for \ion{Mg}{2} absorption systems found
in the survey is shown in Figure \ref{fig_mmtz}.  Also shown is the
sightline coverage for $W_0^{min}$ = 1.0\AA, 0.6\AA, and 0.3\AA.  The
resulting $\partial N/\partial z$ values as a function of look-back
time in a $(\Omega_\lambda,\Omega_M,h)=(0.7,0.3,0.7)$ cosmology are
shown in Figure \ref{fig_mmtdndz}.  The lowest redshift bin
corresponds to the range of look-back time not covered by the EDR
survey, except for the panels with $W_0^{\lambda2796} \ge 2.0$\AA,
which did not have enough absorbers to allow for more than a single
bin.  Also shown are the respective data from the EDR and CRCV99
surveys.  The dashed lines are the no-evolution curves normalized to
the binned EDR/CRCV99 data, {\it excluding} the MMT data.  The $0.367
\le z < 0.956$ MMT results are consistent with the other studies.
Below $z\approx 0.4$, the MMT results for systems with
$W_0^{\lambda2796} \lesssim 0.6$ \AA\ and $W_0^{\lambda2796} \gtrsim
1.5$ \AA\ are consistent within the large uncertainties with no
evolution (however, the EDR results give significant evidence that the
incidence of the strongest systems does indeed decrease relative to
the no-evolution prediction with decreasing redshift).   Also, the MMT
result for 1.0 \AA\ $\le W_0^{\lambda2796} < 1.5$\AA\ at $0.195 < z <
0.367$ is significantly ($\approx 1.5\sigma)$ below the EDR-normalized
no-evolution curve.

NTR05 demonstrated a significant detection of evolution in the
incidence of $W_0^{\lambda2796} \gtrsim 2$\AA\ systems at redshifts $z
\lesssim 1$.  Although our MMT sample is not large enough to confirm
this result for large $W_0^{\lambda2796}$, we do detect evolution at
lower redshift ($0.195 < z < 0.367$) for systems with 1.0 \AA\ $\le
W_0^{\lambda2796} < 1.5$\AA.  Furthermore, the values of $W^*$ derived
from the MMT data (Figure \ref{fig_mmtdwdz}) are consistent with the
NTR05 results whereby the evolution of the incidence of \ion{Mg}{2}
systems is described by a steepening of $\partial N/\partial
W_0^{\lambda2796}$ with decreasing redshift.  While $W^*$ was measured
from the SDSS EDR data down to $\left<z\right>\simeq 0.7$, the MMT
data extend this result down to $\left<z\right>\simeq0.4$.

\section{Discussion}
The incidence $\partial^2 N/\partial W_0 \partial z$ of absorption
systems is proportional to the product of the number density of
systems at a given $W_0$ and redshift times their average absorption
cross section.  For the \ion{Mg}{2} absorbers considered in this work,
$W_0^{\lambda2796}$ is more a measure of kinematics than of column
density, since many of the kinematically-distinct $\lambda2796$ lines
comprising an absorber (which are not individually resolved in the EDR
or MMT data) are typically saturated.  This is particularly the case
for intermediate/strong systems.  We have shown that the total proper
cross section of \ion{Mg}{2} absorbers decreases with cosmic time at a
$W_0^{\lambda2796}$-dependent rate.  While the reason for the
differential evolution is not yet clear, multiple physically-distinct
populations/processes contributing to the incidence of lines
non-uniformly in $W_0^{\lambda2796}$ would provide a explanation if
those populations/processes evolve at different rates.

In NTR05 we discuss evidences for \ion{Mg}{2} absorbers tracing a
variety of physical systems.  However, the possibility that the
observed evolution is to some extent an artifact of an evolving bias
due to dust-induced extinction must also be considered.  We further
discuss these scenarios below.

\subsection{Population-dependent Evolution}
According to Rigby, Charlton \& Churchill (2002), the number of
individual kinematic components (``clouds'') in a given \ion{Mg}{2}
absorber obeys a Poissonian distribution, except for a large excess of
single-cloud systems which account for $\approx 2/3$ of weak  systems
and have their origins in a population of objects and/or processes
distinct from stronger absorbers.  Indeed, an extrapolation of our
exponential fit to $\partial N/\partial W_0^{\lambda2796}$ (\S
\ref{sec_results}) to systems with $W_0^{\lambda2796} \le 0.3$\AA\
accounts for $\approx 30\%$ of the systems with $0.02 \mathrm{\AA} <
W_0^{\lambda2796} < 0.3$\AA\ predicted by the CRCV99 results.  Thus,
it appears that a fraction of the ``weak'' absorbers are physically
similar to intermediate-strength systems, while the majority (the
single-cloud systems) have a different physical nature.  While
intermediate-strength systems show no evolution to within fairly
strong limits over the large range $0.5 \lesssim z \lesssim 2$ (Figure
\ref{fig_mmtdndz}, also see NTR05 for discussion), there is evidence
that the incidence of weak systems exhibits a significant decrease
above $z \approx 1$ (Lynch, Charlton \& Kim, 2005).  It should also be
noted, however, that most single-cloud systems have $W_0^{\lambda2796}
< 0.1$\AA\ (the limit of this study).  Thus, multi-cloud weak systems
may be in part responsible for the upturn between 0.1\AA\ $\le
W_0^{\lambda2796} < 0.3$\AA\ (see also Masiero et al., 2005).  These
systems have been linked in part to dwarf galaxies (e.g., Zonak et
al., 2004).

Imaging studies of intermediate/strong \ion{Mg}{2} absorbers (Rao et
al. 2003, Belfort-Mihalyi et. al, in prep) have shown these systems to
be associated with galaxies that are drawn from a variety of
morphological types.  Also, the results of Kacprzak, Churchill \&
Steidel (2005) indicate that, while galaxy orientation shows no
correlation with absorption properties, interactions are an important
factor in determining the kinematic spread of the absorption in
intermediate strength systems.  Unfortunately, only a handful of
systems have been studied in depth with complementary
imaging/spectroscopy of the absorbing galaxy together with high
resolution spectroscopy to reveal the absorption kinematics.  Steidel
at al. (2002) have shown that extended rotating disks contribute
partially to the incidence in a subset of systems chosen to be
associated with inclined spirals.  However, the edge-on spiral studied
by Ellison, Mall\'{e}n-Ornelas, \& Sawicki (2003) is inconsistent with
extended disk  absorption.  They suggest superbubbles as the cause of
the absorption, as postulated by Bond et al. (2001a).  The few
$W_0^{\lambda2796} \gtrsim 2$\AA\ systems for which high-resolution
spectroscopy exists show kinematics consistent with the superwind
picture (Bond et al., 2001b), though it has also been suggested that
galaxy pairs/interactions may contribute to the incidence of such
systems (see, e.g., Churchill et al., 2000).

Though the strongest systems have been poorly studied due to their
relatively small incidence, there is evidence indicating that the
populations and processes associated with these systems differ from
those associated with intermediate strength absorbers.  Turnshek et
al. (2005, see also Nestor et al. 2003) demonstrate that the average
metallicity in \ion{Mg}{2} absorbers  is strongly correlated with
absorption kinematics, which suggests that the strongest systems are
associated with more massive and/or evolved galaxies.  Furthermore,
M\'{e}nard et al. (in prep) find that the average (over a large number
of absorbers) degree to which QSOs are reddened due to foreground
\ion{Mg}{2} absorbers is strongly $W_0^{\lambda2796}$-dependent for
strong systems.  Scenarios involving both superwinds and interactions
give a natural explanation for the increased metallicity and reddening
with increasingly stronger systems.  Interactions can both strip gas,
causing large line-of-sight velocity spreads (i.e., large
$W_0^{\lambda2796}$ values) and induce star formation which in turn
enriches the gas.  In the superwind picture, relatively highly
enriched gas expelled in superwinds dominates in the strongest
systems, whereas much of the gas responsible for most
intermediate-strength absorbers has remained at large galactocentric
distances (impact parameters) since it first condensed.

As mentioned by Heckman (2002), the appearance of superwinds is
intimately related to the star formation rate (SFR) per unit area in a
galaxy.   While Lyman break galaxies ($z \sim 3$) and
moderate-redshift ($1.4\lesssim z \lesssim 2.5$) star-forming galaxies
do show evidence of superwinds (Pettini et al., 2001; Shapley et al.,
2003; Steidel et al., 2004), the SFR densities of ordinary local
spirals (Kennicutt, 1998) are insufficient.  The incidence of
superwinds should therefore decrease with decreasing redshift,
especially at $z\lesssim1$ since the global SFR decreases below this
value (Hopkins, 2004, and references therein).  The evolution of the
galaxy pair-fraction and merger rates is uncertain, however.  Some
studies have found little evolution over $0 < z \lesssim 1$ (Lin et
al., 2004; Carlberg et al., 2000), while others have found strong
evolution (Le F\`{e}vre et al. 2000).

We have also begun an imaging study of individual ``ultra-strong''
systems, having $W_0^{\lambda2796} \gtrsim 4$\AA.  Though the numbers
are still quite small, early results suggest that ultra-strong
\ion{Mg}{2} absorbers may select galaxies that are preferentially
bright compared to the field population or those selected by
intermediate/strong  \ion{Mg}{2} absorbers (see Nestor, Turnshek \&
Rao, 2005b).

It is unclear, however, how DLA systems fit into the overall picture.
According to Rao, Turnshek \& Nestor (2005) the likelihood of a
\ion{Mg}{2} absorber having a neutral hydrogen column density above
the DLA threshold ($N_{HI} \ge 2\times10^{20}$) rises from $\approx
15\%$ for $W_0^{\lambda2796} \simeq 0.6$\AA\ to $\approx 65\%$ for
$W_0^{\lambda2796} \simeq 3.0$\AA.  However, DLAs preferentially
select neither bright galaxies, as the ``ultra-strong'' systems may,
nor high-SFR systems (see Hopkins, Rao \& Turnshek, 2005), which are
required for superwinds.  If the incidence of the DLA, superwind, and
interacting/galaxy-pair populations are evolving at a different rate
than that for the population/processes traced (at least in part) by
the ``classic'' \ion{Mg}{2} absorbers, this should contribute to the
differential evolution seen in \ion{Mg}{2} $\partial N/\partial z$.
As the role of the DLA sub-population provides important clues for the
study of low-ion/neutral gas, sorting this out should be an important
goal of future work.

\subsection{Dust}
The differential evolution in $\partial N/\partial z$ could also be
affected by an evolving dust bias.  Vladilo et al. (2005) claim that
extinction from dust increases linearly with Zn column  density, from
small ($A_V \lesssim 0.02$ mag) levels at $N_{Zn} \lesssim 10^{12}$ to
significant levels at higher column densities ($A_V \approx 0.15$ mag
at $N_{Zn} \approx 10^{12.8}$).  The results of Turnshek et al. (2005)
indicate that the average Zn column density increases to approximately
this level for strong ($W_0^{\lambda2796}\approx 3$\AA) \ion{Mg}{2}
systems.\footnote{Note, that since these are averages, some systems
are expected to have larger Zn columns, even at smaller
$W_0^{\lambda2796}$ values.}  Indeed, the M\'{e}nard et al (in prep)
results for E(B-V) versus $W_0^{\lambda2796}$ are roughly consistent
with that predicted by Vladilo et al., considering the Turnshek et
al. $W_0^{\lambda2796}$--$N_{Zn}$ relation.  Thus, the largest
$N_{Zn}$ systems (which preferentially correspond to the largest
$W_0^{\lambda2796}$ values) maybe be missed due to the background QSOs
dropping out of magnitude limited and/or color selected surveys such
as the SDSS.  If this is indeed the case, we expect that the
$W_0^{\lambda2796}$--$N_{Zn}$ relation will level off at large
$W_0^{\lambda2796}$ values as the largest individual $N_{Zn}$ systems
drop out of the survey.  Also, we would expect a turnover in $\partial
N/\partial W_0^{\lambda2796}$ at large  $W_0^{\lambda2796}$ as these
systems are preferentially missed.\footnote{However, the presence of a
turnover, if found, could also be explained by a physical upper-limit
to the velocity spread of \ion{Mg}{2} absorbers.}  Both of these tests
will be possible with larger \ion{Mg}{2} absorber catalogs formed from
the full SDSS survey.  We also note that although recent studies such
as those emerging from the CORALS survey (e.g., Ellison et al., 2004)
claim to find no dust biases, the size of their survey is not large
enough to study the rare, strongest systems.

\section{Conclusions}
We have detected and measured 140 \ion{Mg}{2} absorption systems in
the spectra of 381 QSOs obtained at the MMT Observatory.  These data
include systems with $\lambda2796$ rest equivalent width in the  range
0.1 \AA\ $\le W_0^{\lambda2796} \le 3.2$ and extend our determination
of the incidence of \ion{Mg}{2} absorbers down to redshift $z=0.15$.

Coverage of $W_0^{\lambda2796}$ across the value
$W_0^{\lambda2796}\simeq 0.3$\AA\ is important since the distribution
$\partial N/\partial W_0^{\lambda2796}$ is represented very well by a
single exponential function above $W_0^{\lambda2796}\simeq0.3$ \AA,
but diverges from the fit with an excess of absorbers below this
value.  This effect was first noted in NTR05 by comparing our SDSS EDR
survey data for $W_0^{\lambda2796}\ge 0.3$\AA\ with data from CRCV99
having $W_0^{\lambda2796} < 0.3$\AA.  Here, we confirm the nature of
the $\partial N/\partial W_0^{\lambda2796}$ distribution in a single
survey thereby removing the concern that the putative upturn was an
artifact of unknown systematic differences in the two independent
surveys.

By extending the redshift coverage of our \ion{Mg}{2} survey to lower
redshift, we sample systems down to a look-back time of only $\approx
2$ Gyrs, compared to $\approx 4$ Gyrs for SDSS samples, and thereby
increase the total look-back time covered by $\approx 1/3$.  The
low-redshift regime is of particular importance, especially for
intermediate/strong systems, since the apparent evolution in their
incidence is easily manifested only at low redshift.  While NTR05
report evolution in systems with $W_0^{\lambda2796}\gtrsim 2.0$\AA\ at
redshifts $z \lesssim 1$, we now detect evidence for evolution in
systems with $W_0^{\lambda2796}\gtrsim 1.0$\AA\ at redshifts $z
\lesssim 0.5$.  The evolution apparent in our MMT sample is consistent
with the nature of the evolution described in NTR05, whereby the
incidence decreases from the no-evolution prediction (i.e., the total
proper absorption cross section decreases) at a rate dependent on
$W_0^{\lambda2796}$.

It seems likely that multiple populations and processes contribute to
the total cross section of \ion{Mg}{2} absorption in a
$W_0^{\lambda2796}$-dependent manner, and that the different
evolutionary rates of these populations and processes lead to the
$W_0^{\lambda2796}$-dependent evolution in $\partial N/\partial z$.
It is also possible, however, that biases due to dust-induced
extinction give rise to the detected evolution.  It will be necessary
to determine the relative contribution of the various
physically-distinct components/processes to the incidence of
\ion{Mg}{2} systems, as well as obtain stronger constraints on the
evolution of the incidence, in order to better understand
low-ion/neutral gas absorption systems.  Determining the relative
contribution as a function of $W_0^{\lambda2796}$ will help us further
understand the evolution in time of each type (and, perhaps,
vice-versa).  The results of our ongoing studies, such as the imaging
of ``ultra-strong'' systems and the analysis of the full SDSS sample,
are two such projects that should add to our understanding of the
nature and evolution of these systems.

\acknowledgements  We acknowledge support for this work from the NSF.
We also acknowledge Brice M\'{e}nard for helpful discussions and
ongoing collaborations related to this work, and thank the MMT staff
and the members of the SDSS collaboration who made our observing
runs and the SDSS project a success.

\end{document}